\documentclass[aps,prl,twocolumn,groupedaddress,showpacs,superscriptaddress,amssymb,amsmath]{revtex4-1}
\usepackage{graphicx}
\usepackage{tabularx}
\usepackage{color}
\usepackage{amsmath}
\usepackage{comment}
\usepackage{amsfonts,amssymb}
\newcommand{\be}{\begin{equation}}
\newcommand{\ee}{\end{equation}}
\newcommand{\bea}{\begin{eqnarray}}
\newcommand{\eea}{\end{eqnarray}}
\usepackage{dcolumn}
\usepackage{bm}
\usepackage{epsf}
\usepackage{subfigure}
\usepackage{epstopdf}%
\usepackage{amsfonts}
\usepackage{hyperref}

\begin{document}
%HARE KRISHNA

%\title{Tuning Optical Properties of a Cavity-Coupled Double Quantum Dot at Finite Bias: Diagrammatic NEGF Approach}
%\title{Tunable Cavity-Coupled Double Quantum Dot under voltage bias: Diagrammatic NEGF Approach}
%\title{Tunable Out-of-Equilibrium Cavity-Coupled Double Quantum Dot: Diagrammatic NEGF Approach}
%\title{Out-of-Equilibrium Cavity-Coupled Double Quantum Dot: Diagrammatic NEGF Approach}
% DDD: why 'optical'? - We think here more concretely about the circuit QED.

%\title{Giant optical amplification in large-scale quantum dot circuit-QED systems}
%DVIRA: how about:
\title{Giant photon gain in large-scale quantum dot circuit-QED systems}
% Giant amplification of photon gain in large-scale   quantum dot circuit-QED systems
% Giant amplification of photon transmission in large-scale   quantum dot circuit-QED systems

\author{Bijay Kumar Agarwalla}
\affiliation{Chemical Physics Theory Group, Department of Chemistry,
and Centre for Quantum Information and Quantum Control,
University of Toronto, 80 Saint George St., Toronto, Ontario, Canada M5S 3H6}

\author{Manas Kulkarni}
\affiliation{Department of Physics, New York City College of Technology, The City University of New York, Brooklyn, New York 11201, USA}

\author{Shaul Mukamel}
\affiliation{Department of Chemistry, University of California, Irvine, California 92697, USA}

\author{Dvira Segal}
\affiliation{Chemical Physics Theory Group, Department of Chemistry,
and Centre for Quantum Information and Quantum Control,
University of Toronto, 80 Saint George St., Toronto, Ontario, Canada M5S 3H6}

\date{\today}

\begin{abstract}
Motivated by recent experiments on the generation of coherent light in engineered hybrid quantum systems,
we investigate gain in a microwave photonic cavity coupled to quantum dot structures, % serving as the gain medium.
and develop concrete directions for achieving a giant amplification in photon transmission.
We propose two architectures for scaling up the electronic gain medium:
(i) $N$ double quantum dot systems (N-DQD), % with each DQD maintained at a finite DC bias,
(ii) $M$ quantum dots arranged in series
%and coupled to a cavity
akin to a quantum cascade laser setup.
In both setups, the fermionic reservoirs are voltage biased, and the quantum dots are coupled to a single-mode cavity.
Optical amplification is explained based on a sum rule for the transmission function, and it is
determined by an intricate competition between two different processes: charge density response in the gain medium, % correlations?
and cavity losses to input and output ports.
The same design principle
is also responsible for the corresponding giant amplification in other photonic observables,
mean photon number and emission spectrum, thereby realizing a quantum device that behaves as a giant microwave amplifier.
%
%DVIRA: optimal: it is not clear what happends when we go beyond N=4. We may get better results, but the theory is not applicable there.
%optimal means that 4 is better than 5. I am not sure about it.
\end{abstract}
\vspace{-0.5cm}
\maketitle

%==============================================
{\it Introduction.-}
\label{intro}
% motivation: hybrid devices
Remarkable progress has been made in engineering, probing,
and controlling hybrid light-matter systems which sit at the confluence of quantum optics and condensed matter
physics \cite{RevNori,nori_sim, PNAS1,Hincks_control,KLH2015,nori_simulator}.
%This progress has been paramount for the advancmenet
%of quantum information and simulation schemes, the generation of coherent light and studies of fundamental nonequilibrium phenomena.
%
Important examples include cavity-quantum electrodynamics arrays
\cite{nat_phys_rev,underwood_PhysRevA.86.023837,sskj}, trapped cold atoms coupled to
photon degrees of freedom \cite{esslinger10, kulkarniprl,Krinner2015,ess_science}, interconnected copper waveguide cavities,
each housing a qubit \cite{kulkarniprx, kulberkeley,kulpra}.
%
%From a fundamental aspect,
%fascinating phenomenon in light-matter systems include open quatum phase transitions \cite{kulkarniprl},
%quantum entanglement\cite{kulkarniprx}, realizations of non-local interacting models \cite{Mottl1570}, non-Markovian dynamics
%to name a few. From the application perspective \cite{RevNori},
%these light matter systems are crucial for realization of novel quantum technologies, simulators \cite{nori_sim} and devices.
%
%
The successful integration of biased quantum dots (mesocopic electronic systems)
with a transmission line resonator (photonic degrees of freedom) % acting as a microwave cavity
has been a major step forward in this field %of hybrid quantum systems
\cite{kontoskondo, kontosnatcom, Viennot2013, Petersson2012, petta2014,nori1, Frey2012,enslinprl,Toida2013,Kulkarni2014,Deng2013,kontosprx,gpg, Vavilov-fcs, Schon_noise}.
Such  quantum dot circuit-quantum-electrodynamics (QD-cQED) hybrids
%achieving sufficiently strong charge-cavity coupling between electronic and photonic degrees of freedom
open up new directions for realzing quantum computing schemes
based on localized electronic spins \cite{spin-kontos, Nori-bistable}, controlling electronic current via `light' \cite{ss1,ss2, ssn1, tb1}, and achieving high gain in the cavity transmission \cite{RevNori,KLH2015}.
Fundamentally, QD-cQED systems serve as a versatile platform for probing non-equilibrium open many-body quantum systems, by realizing
basic models and phenomena in physics, for e.g.,
the Anderson-Holstein Hamiltonian with the fermionic system
tuned to the Coulomb blockade or the Kondo regime \cite{kontoskondo, kontosprx}.
%As well, deviations from certain universal results, e.g., the
%Korringa-Shiba relation, can be probed by tuning
%the electronic system \cite{kontosprx}.
%
%================================
% Figure 1
\begin{figure}[h!]
\includegraphics[width=8.2cm]{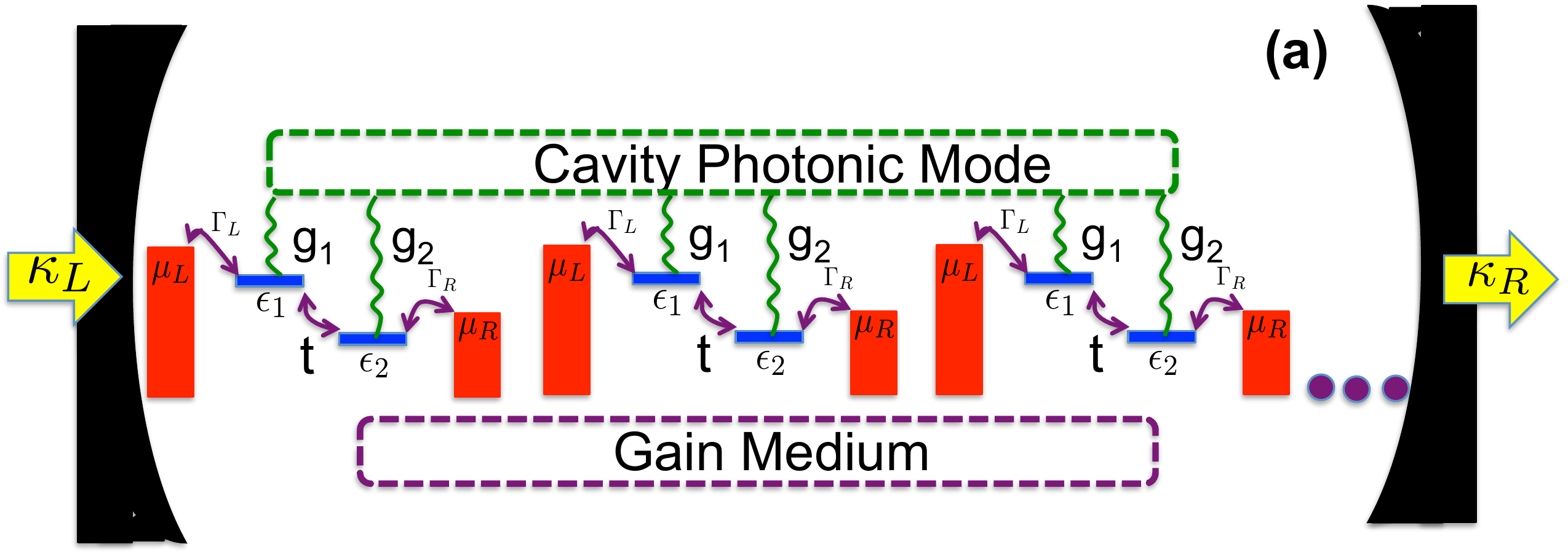}
\\ 
\includegraphics[width=8.2cm]{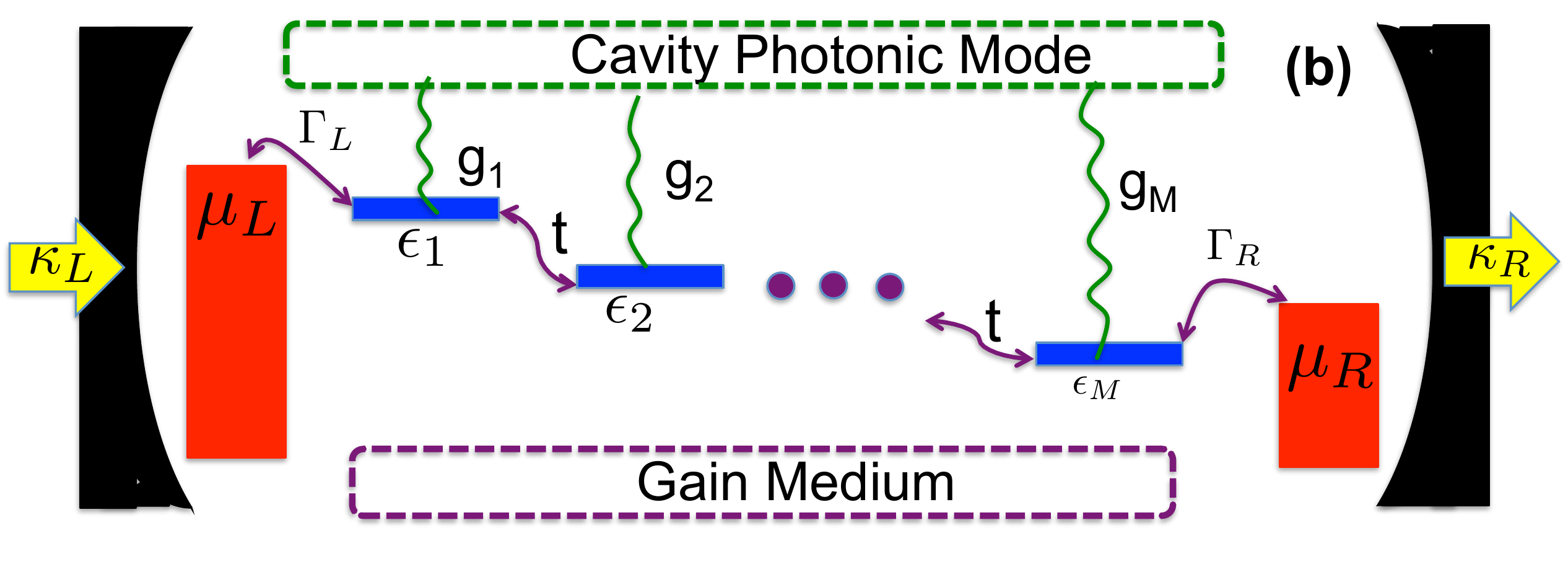}
\caption{(Color online)
Schemes of large-scale quantum dot circuit-QED systems designed for achieving giant optical gain.
A transmission line resonator is coupled via a Holstein-like interaction to an electronic gain medium with
(a) $N$ double quantum dots, each DQD
tunnel-coupled to external electrodes and driven out-of-equilibrium by the application of a source-drain bias.
%$\Delta \mu =\mu_L-\mu_R$.
(b) $M$ dots in a cascade setup, with the first and last sites coupled to electronic leads.
%
%In both models
Tunnelling rates between the dots and the electron leads ($\Gamma_L , \Gamma_R$) and in between the dots $(t)$
are tuned via gate-controlled tunnel-barriers. Cavity photons are coupled to the input and output ports with rates $\kappa_{L(R)}$.
Arrows represent tunnelling processes and wavy lines indicate light-matter couplings.
}
\label{model-scheme}
\end{figure}
%=============================
%

Focusing on the optical properties of the cavity,
%as inelastic charge transport through the quantum dots occurs via the absorbtion or emission of microwave photons
%in the resonator,
QD-cQED  devices can be engineered and optimized to increase photon emission, by
utilizing the voltage biased QDs as a gain medium \cite{petta2014,enslinprl}.
To significantly enhance optical signal in the cavity, %  and ultimately demonstrate the action of a maser,
recent efforts were focused on {\it scaling up} the gain-medium \cite{tk2011,Schon}. % Add more exp  
A major advance in this regard has been the realization of a microwave laser  (\textit{maser})
via the fabrication of a {\it double} double quantum dot gain medium. 
%two voltage-biased DQDs were
%individulally coupled to the same cavity mode, and
In this setup, only when {\it both} electronic units were properly tuned to the cavity frequency did a maser action appear \cite{pettamaserscience}.
%The role of a gain medium was successfully enhanced to lead to a maser phenomenon. These experiments in QD-cQED systems are versatile, tunable and importantly, scalable.
Despite impressive experimental demonstrations,
a theoretical understanding of principles governing amplification of photon emission in hybrid light-matter devices
%cavity-coupled quantum dot devices
is missing. Specifically, what architectures, comprising
large scale electronic quantum dot systems, can act as momentous gain media? 
How should  we tune together the different couplings and driving forces %in the system 
%electron-photon interaction, electron dissipation, photon loss, voltage bias, 
to realize a giant microwave amplifier?

% Here

In this Letter, we describe from microscopic principles directions for enhancing photon emission, eventually reaching the lasing
threshold---from below. We achieve a giant amplification of photonic observables (transmission, mean photon number, emission spectrum)  by employing different large-scale gain media,
(i) an $N$ double quantum dot system (N-DQD) with each DQD maintained at a finite dc bias, see Fig \ref{model-scheme}(a),
(ii) $M$ dc-biased quantum dots arranged in series, see Fig \ref{model-scheme}(b).
The second scenario is similar in spirit to the quantum cascade laser setup, and we refer to it as the Quantum Cascade (QC) model \cite{qcl}.
In both cases the electronic systems are coupled to the cavity with a \textit{Holstein-like} light-matter interaction model.
For the N-DQD system, a simple scaling law for the gain medium is identified,
to reach giant amplification of photonic properties, significantly larger than a naive sum of individual gains for each DQD. % combinataion - is it product? (Science paper)
In contrast, the QC device is missing such scalibility, yet we can identify cases beneficial for gain.
%but displays a fascinating interplay between optical and electron properties and thereby shows interesting open quantum many body phenomena.% DVIRA: ... I know it is a matter of style, but this sentence is empty. Here we should put concrete information.
%

{\it N-DQD Gain Medium.-}
\label{model}
We begin with the model displayed in Fig.~\ref{model-scheme}(a) 
to explain the mechanism of photon amplification in the cavity.
The electronic gain medium consists of $N$ DQDs coupled to the same microwave cavity.
Each DQD is further coupled to electronic leads at finite bias, denoted by $\Delta \mu =\mu_L-\mu_R$.
%which drives the system away from equilibrium. DVIRA: obviously.
The total Hamiltonian consists the $N$ fermionic replicas,
$\hat H= \sum_{i=1}^{N}\hat H_{el}^{i} + \sum_{i=1}^{N}\hat H_{el-ph}^{i} + \hat H_{ph}$,
with $ \hat H_{el}^{i} = \hat H_{QD}^i +\hat  H_{lead}^{i} + \hat H_{T}^{i}$.
Here $\hat H_{QD}^i= {\epsilon} \,(\hat n_{i1}-\hat n_{i2})/2 + t ( \hat d_{i1}^{\dagger} \hat d_{i2} + h.c.)$ is the $i$-th
DQD hamiltonian with
$\epsilon$ as the detuning energy, $t$ the hopping parameter, and  $\hat n_{i1,i2} = \hat d_{i1,i2}^{\dagger} \hat d_{i1,i2}$ are the
number operator for dots 1,2 respectively. 
%in the i-th DQD.} 
Each DQD is connected to two electronic leads $\alpha=L,R$,
$\hat H_{lead}^i= \sum_{k, \alpha} \epsilon_{k \alpha} \hat c_{i k \alpha }^{\dagger} \hat c_{i k \alpha }$, where $k$ is the index for momentum, 
with the standard tunnelling
Hamiltonian $\hat H_{T}^i = \sum_{k} v_{i  k L} \hat d_{i1}^{\dagger} \hat c_{i k L} +
\sum_k v_{i k R} \hat d_{i2}^{\dagger} \hat c_{i k R} +
 h.c$. $\hat d$ and $\hat c$ are fermionic annihilation operators, $h.c$ denotes Hermitian conjugate. We define the spectral density for the electronic leads as 
$\Gamma_{\alpha}^{i}(\omega) = 2 \pi \sum_{k} |v_{i k \alpha} |^2 \delta(\omega-\epsilon_{k\alpha})$.  
The photonic (bosonic) hamiltonian $\hat H_{ph}$ consists of the cavity mode of frequency $\omega_c$, and two long transmission lines,
left and right, ($K=L,R$) with coupling $\nu_j$ to the cavity mode,
$\hat H_{ph} = \omega_c  \hat a^{\dagger} \hat a +
\sum_{j\in K} \omega_{jK} \hat a_{jK}^{\dagger} \hat a_{jK} + \sum_{j\in K} \nu_{j} \,\hat a_{jK}^{\dagger} \, \hat {a} + h.c.$
The interaction between the microwave photon and the dipole moment of excess electrons in the DQDs
is given by $ \hat H_{el-ph}^i =g_i\, (\hat{n}_{i1}- \hat{n}_{i2}) (\hat a^{\dagger} + \hat a)$, with $g_i$ as the coupling
strength between the $i$-th DQD and the cavity.
In what follows we assume that the DQDs replicas are identical, thus ignore the index $i$ when appropriate.

We investigate the cavity response by focusing on the transmission function. Experimentally, such measurements are performed via heterodyne detection which can be realized here by interpreting the bosonic modes of the left and the right transmission lines as the input and output microwave signals, respectively. Following the input-output theory \cite{input_output_Clerk,marco14,simon16},
the transmission function $t(\omega)$ (ratio of output vs input signal) for a single DQD $(N\!=\!1)$
can be expressed in terms of the response function of the cavity mode as
\be
t(\omega) \!=\! \frac{i \kappa}{(\omega-\omega_c) + i \kappa -F_{el}^r(\omega)} \!=\!i \, \kappa \,D^r(\omega),
\label{eq:tdef}
\ee
where $D^r(t)= -i \,\theta(t) \, \langle [\hat{a}(t),\hat{a}^{\dagger}(0)]\rangle$ is the response function with the
average performed over the combined electronic and photonic degrees of freedom. We further identify the electronic charge susceptibility in the time domain by 
$F_{el}^{r}(t-t')= g^2 \, \sum_{l, j =1,2} (2 \,\delta_{l j} -1 )\, \Lambda^{el}_{lj}(t-t')$. Here
\be
\Lambda^{el}_{lj}(t-t')= -i \, \theta(t-t') \, \big\langle \big[\hat{n}_{l}(t), \hat{n}_{j}(t')\big]\big\rangle_{el(g=0)}
\ee
is the electron density response function, with the average performed over the electronic medium (dots and leads). In Eq.~(\ref{eq:tdef}), $\kappa\!=\!\kappa_L\!=\!\kappa_R$ is the decay rate of the cavity mode per port \cite{comment1}.
Experimentally, it is large compared to  $|F_{el}^r(\omega)|$ \cite{petta2014, pettamaserscience}.

Inspecting Eq.~(\ref{eq:tdef}), we immediately identify a simple-fundamental principle for achieving gain, $|t(\omega)|^2>1$: We need to counteract
the two different sources of response, %dissipation,
the imaginary component of the gain medium-induced self-energy  $F_{el}^{''}(\omega)\equiv {\rm Im}[F_{el}^r(\omega)]$,
and the cavity decay rate to the ports.
In other words, $F_{el}^{''}(\omega)$ should approach $\kappa$ for achieving maximum gain.
This objective cannot be accomplished at equilibrium,  as $F_{el}^{''}(\omega)<0$ \cite{simon16}. Therefore, driving the electronic system out-of-equilibrium is a necessary condition for gain.
Most significantly, from the causality condition of the retarded Green's function we receive the sum rule
\be
\int_{-\infty}^{\infty} \frac{d\omega}{2\pi} \, t(\omega)= \frac{\kappa}{2},
\ee
valid  for $\kappa > F_{el}^{''}(\omega)$. It tells us that an enhancement in maximum gain must be accompanied with the reduction in the width of the emission spectrum, thereby increasing the coherence time significantly, a critical requirement to eventually realize a maser \cite{pettamaserscience}.
%This relation follows from the causality condition of the retarded Green's function.
Explicitly, $\int_{-\infty}^{\infty} \!\frac{d\omega}{2\pi} \,{\rm Re}[t(\omega)]\!=\!\kappa/2$ and $\int_{-\infty}^{\infty} \!\frac{d\omega}{2\pi} \,{\rm Im}[t(\omega)]\!=\! 0$.
% The sum rule for ${\rm Re}[t(\omega)]$ indicates that for a fixed $\kappa$,
%an increase in the maximum transmission shows up with the reduction of the width of the spectrum,
%thereby increasing the coherence time significantly, a critical requirement to eventually realize a maser \cite{pettamaserscience}.
%

\begin{figure}
\includegraphics[scale=0.3]{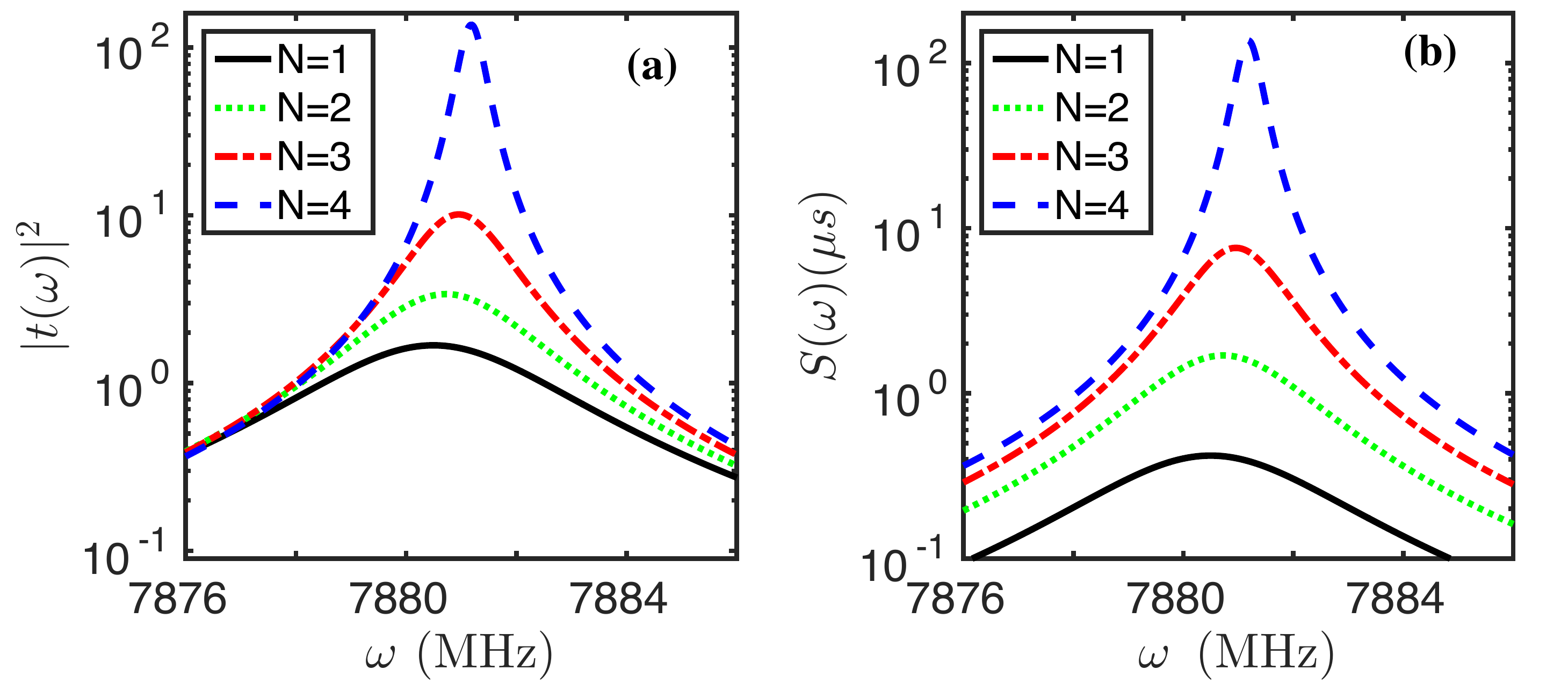}
\caption{(Color online) (a) Gain $|t(\omega)|^2$ and (b) emission spectrum $S(\omega)$ as a function of incoming frequencies $\omega$
for different number of DQDs. Parameters are $g= 50$ MHz, $\kappa=3.15$ MHz, $\omega_c=7880.5$ MHz,
$\Gamma=2.6, \epsilon=7.0, t=16.4, \Delta \mu=250$, $k_BT=0.69$, all in $\mu$eV.}
\label{trans_omega}
\end{figure}

Our objective is to enhance the electronic response $F_{el}^{''}(\omega)$ to reach high gain even for a poor (lossy) cavity with high rate $\kappa$. It can be optimized to a certain extent
in a single DQD by tuning the metal-dot hybridization $\Gamma$ and the bias voltage.
%which may eventually lead to an experimental realization of the analog of a single-atom maser. %, as we show below.
We suggest an alternative, simple yet powerful, scalable approach: include $N$ replicas of the DQD system
to extensively-linearly increase the self-energy $F_{el}^r(\omega)$ \cite{comment-coup}.
For the case of $N$ DQDs, the absolute value of the transmission, defined via $t(\omega)=|t(\omega)| e^{i \phi(\omega)}$, is now given as
\be
|t(\omega)|^2= \frac{\kappa^2}{\big[\omega\!-\!\omega_c\!-\!N\,F_{el}^{'}(\omega)\big]^2 + \big[\kappa \!-\! N\,F_{el}^{''}(\omega)\big]^2},
\label{trans-final}
\ee
with the area law $\int_{-\infty}^{\infty} \frac{d\omega}{2\pi}\, |t(\omega)|^2 = \frac{\kappa^2}{2\, \big[\kappa \!-\! N\,F_{el}^{''}(\omega_c)\big]}.$
Here, $F_{el}^{'} \,(F_{el}^{''})$ stands for real (imaginary) component of $F_{el}^r(\omega)$. It is clear that 
the transmission peak shifts from $\omega_c$ by $N F_{el}^{'}(\omega_c)$, and the peak value is determined by the
the difference between the electronic response and photon loss, $\kappa-N\, F_{el}^{''}(\omega_c)$. 
Fig.~\ref{trans_omega} demonstrates this enhancement mechanism for a fixed detuning $\epsilon$.
With increasing number of DQDs, the transmission shows significant gain, as well as a reduction in width---close to the cavity frequency $\omega_c$.
%The peak value is blue shifted compared to the zero current case.
In our parameters, $F_{el}^{''}(\omega_c)$ approaches $\kappa$ for $N_c=4$, materializing giant gain.
The detuning $\epsilon$ was chosen to satisfy a resonant condition, $\omega_c\sim \sqrt{\epsilon^2+4t^2}$.
We later show that by searching for an optimal $\epsilon$ one can enhance the maximum gain by five orders of magnitude relative to the $N$=1 case.

Another relevant measure for the cavity response is the emission spectrum, induced by the electronic current,
defined as $S(\omega)= \int_{-\infty}^{\infty} dt \langle \hat{a}^{\dagger}(0) \hat{a}(t) \rangle e^{i \omega t} = i \, D^{<}(\omega)$.
It takes a structure similar to Eq.~(\ref{trans-final}),
\be
S(\omega)=i \frac{N \, F_{el}^{<}(\omega)}{\big[\omega\!-\!\omega_c \!-\!N \,F^{'}_{el}(\omega)\big]^2\!+\! \big[\kappa \!-\! N\,F_{el}^{''}(\omega)\big]^2},
\ee
and hence it can be similarly amplified, see Fig. \ref{trans_omega}(b).
It immediately follows that $\int_{-\infty}^{\infty} \frac{d\omega}{2\pi} S(\omega) = \langle \hat{a}^{\dagger} \hat{a} \rangle \equiv \langle \hat{n}_c \rangle$.

%===================================
% Figure 2

% Figure 3
\begin{figure}
\includegraphics[scale=0.28]{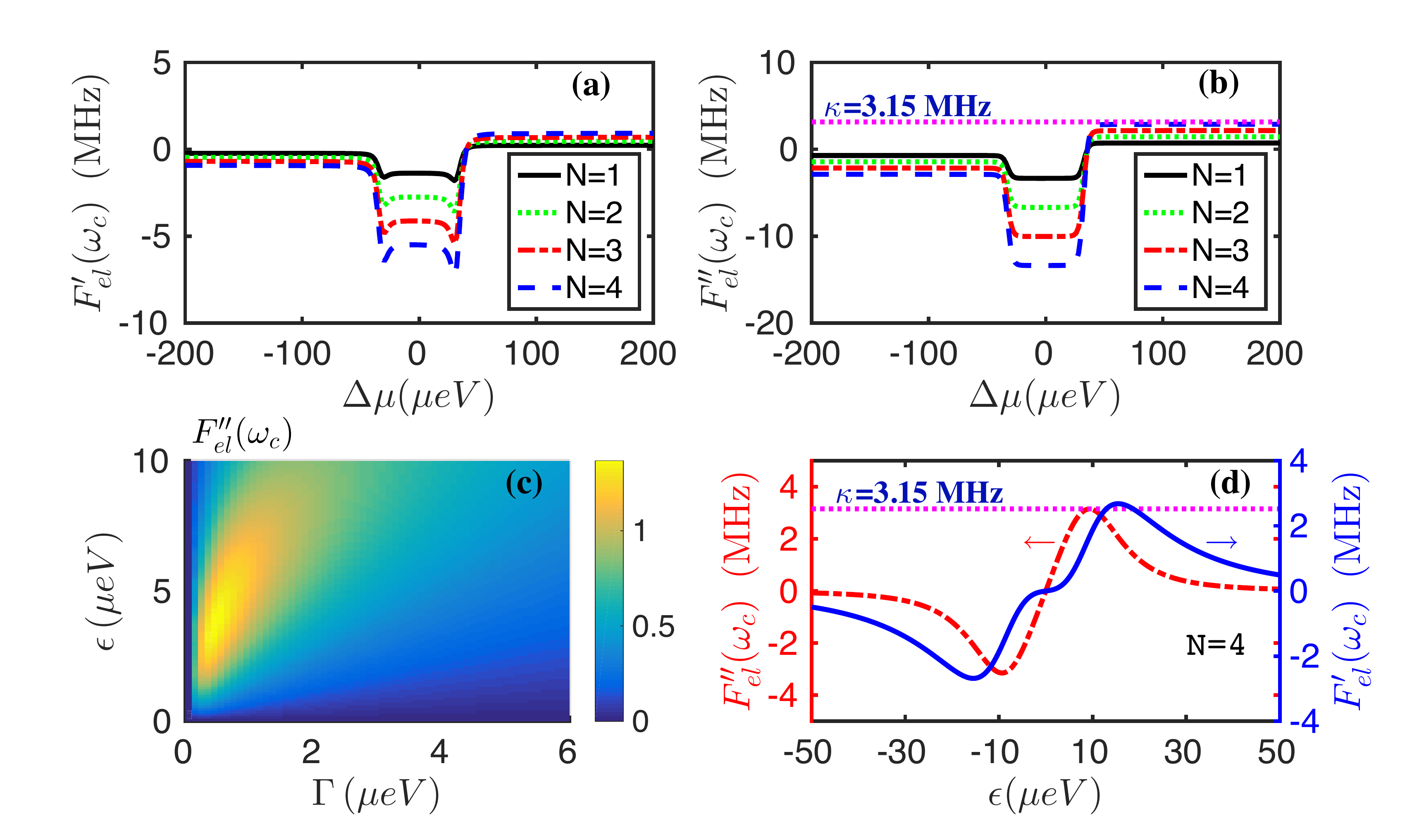}
\caption{(Color online) (a) Real ($F_{el}^{'}(\omega_c))$ and (b) imaginary components ($F_{el}^{''}(\omega_c))$ of $F_{el}^r(\omega)$ vs.
bias difference $\Delta \mu$.
(c) Two dimensional plot of $F_{el}^{''}(\omega_c)$ as a function of dot-lead coupling $\Gamma$ and detuning $\epsilon$.
(d) $F_{el}^{'/''}(\omega_c)$ vs. $\epsilon$ for $N\!=\!4$.
Other parameters are same as in Fig.~(\ref{trans_omega}).}
\label{trans_detu}
\end{figure}
%=================================

%=============================================================
% Figure 4
\begin{figure}[t]
\includegraphics[scale=0.3]{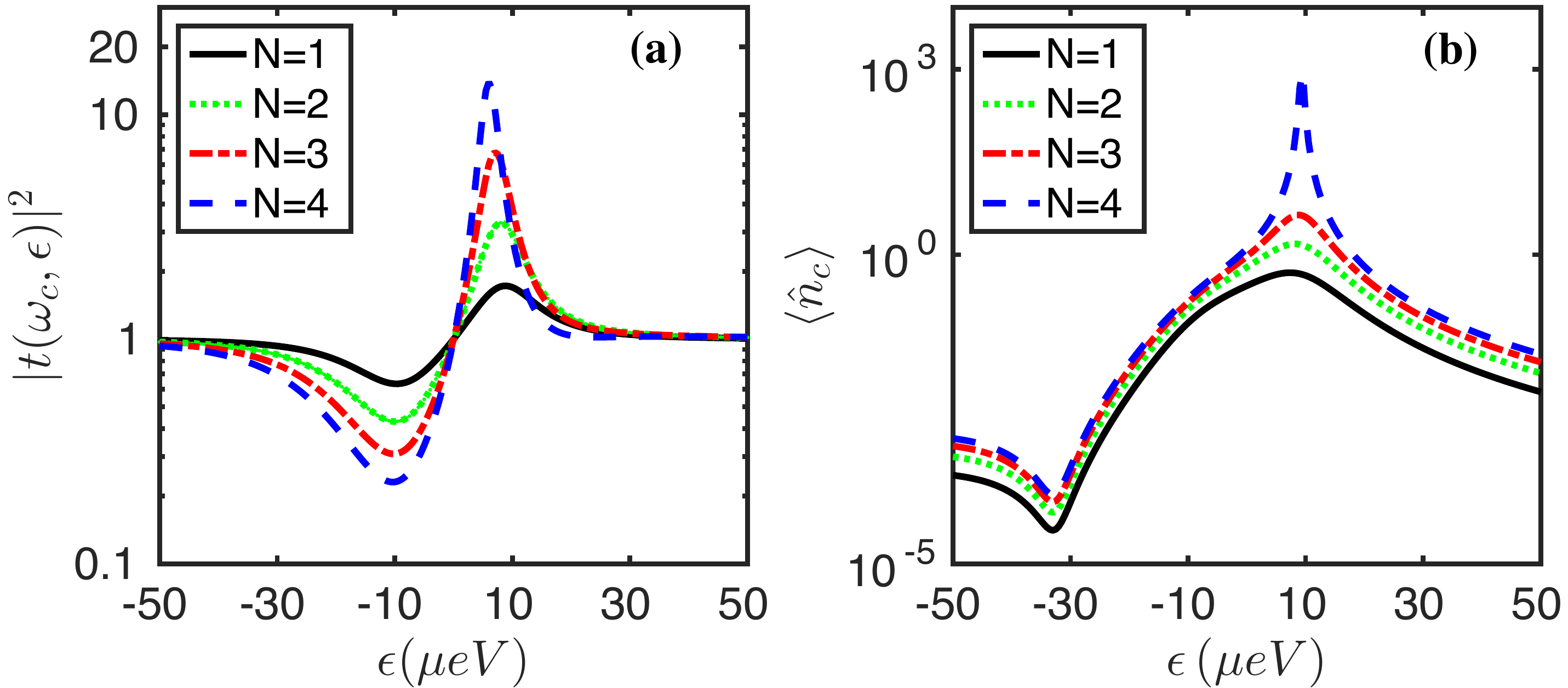}
\caption{(a) Gain $|t(\omega_c, \epsilon)|^2$ and (b) average photon number $\langle \hat{n}_c \rangle$ as a function of energy detuning $\epsilon$ for different number of DQDs. Other parameters are same as in Fig. \ref{trans_omega}.}
\label{avg_photon}
\end{figure}
% Figure 5
\begin{figure*}[ht!]
\includegraphics[scale=0.4]{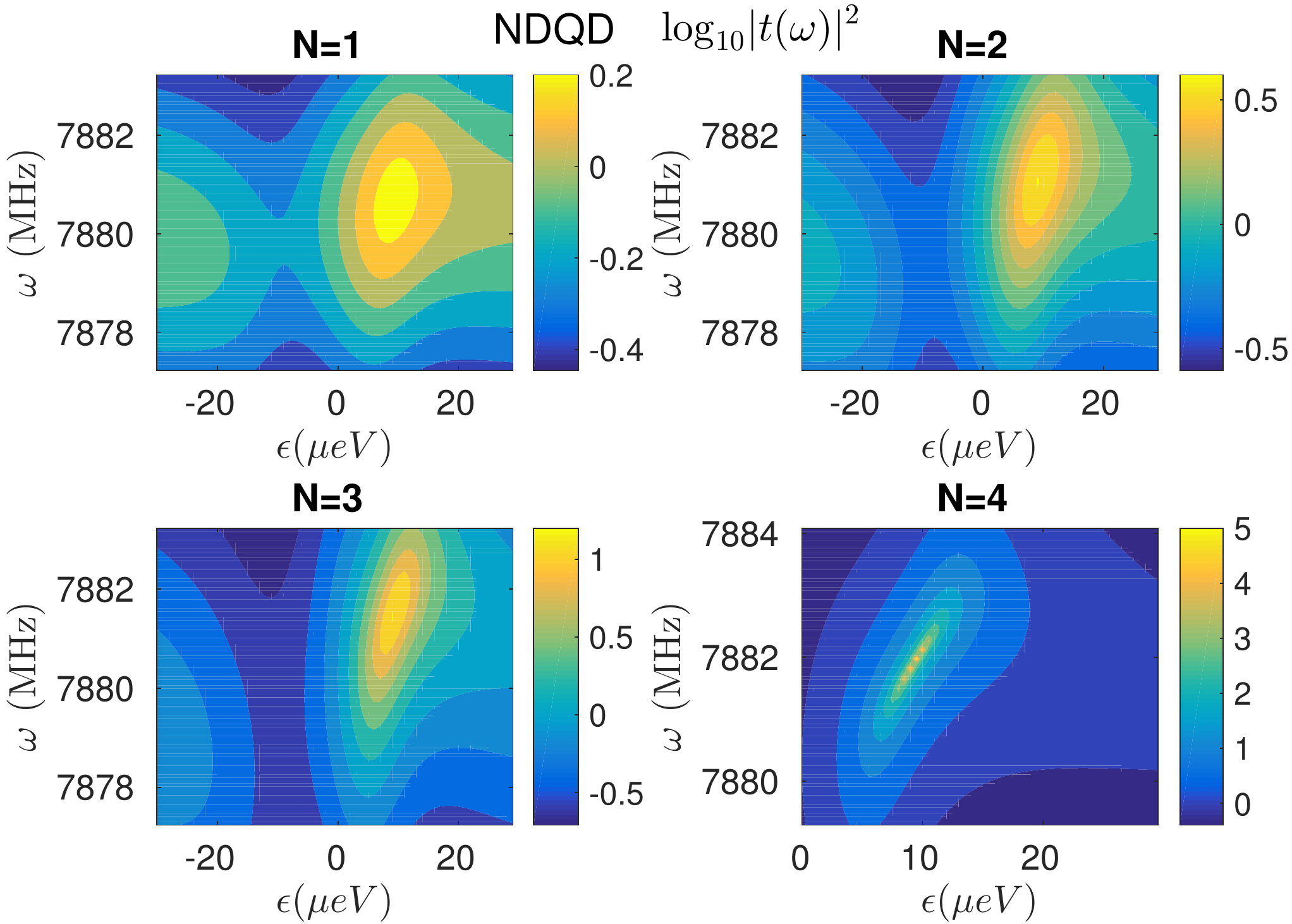}
\hfill
\includegraphics[scale=0.4]{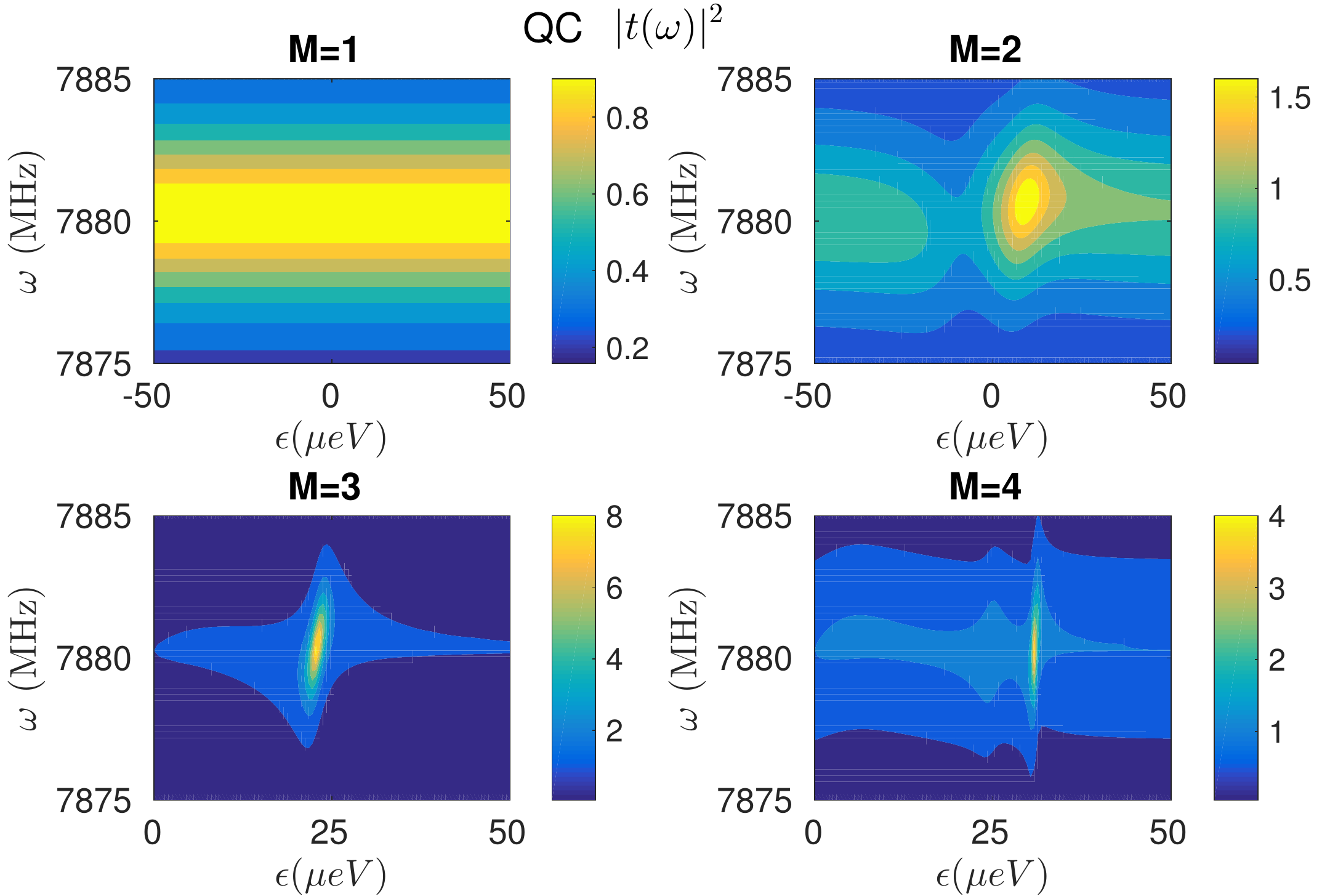}
\caption{Two dimensional plot of $|t(\omega)|^2$ as a function of $\omega$ and detuning $\epsilon$ for the N-DQD model
(left $2\times 2$ panels, also note the $log_{10}$ scale)
and the QC model (right $2\times 2$ panels).
Parameters are same as in Fig. \ref{trans_omega}.}
\centering
\label{cascade}
\end{figure*}

Explicit expressions for the different components of the self energy, $F^{','',<,>}_{el}(\omega)$,
can be derived by employing a scheme based on
the random-phase approximation, which is correct up to the second order of light-matter coupling but non-perturbative in the dot-lead coupling.
With the help of Keldysh NEGF technique and the Langreth formulae \cite{NEGF-Rammer}, we receive  % formula or formulae
the real and imaginary components of the self energy as \cite{longNEGF}
\bea
F_{el}^{'}(\omega) \!&=&\! -i\int_{-\infty}^{\infty}\!
\!\frac{d\omega'}{8\pi} \Big\{ {\rm Tr} \Big[{\bf g} \, {\bf
G}_0^{k}(\omega_+)\, {\bf g}\, \big({\bf G}_{0}^{r}(\omega_-)\!+\! {\bf
G}_{0}^{a}(\omega_-)\big)\Big]
\nonumber \\
&&\!+\! {\rm Tr}\Big[ {\bf g}\,{\bf
G}_{0}^{k}(\omega_-)\, {\bf g}\, \big({\bf G}_{0}^{r}(\omega_+)\!+\! {\bf
G}_{0}^{a}(\omega_+)\big) \Big]\Big\}, 
\nonumber\\
F_{el}^{''}(\omega) &=& \int_{-\infty}^{\infty} \frac{d\omega'}{8\pi}
\Big\{ {\rm Tr} \Big[{\bf g} \, {\bf G}_0^{k}(\omega_+)\, {\bf g}\,
\big({\bf G}_{0}^{r}(\omega_-)\!-\!{\bf G}_{0}^{a}(\omega_-)\big)\Big] \nonumber \\
&-&{\rm
Tr}\Big[ {\bf g}\,{\bf G}_{0}^{k}(\omega_-)\, {\bf g}\, \big({\bf
G}_{0}^{r}(\omega_+)\!-\!{\bf G}_{0}^{a}(\omega_+)\big)\Big]\Big\},
\label{eq:real-im}
\eea
which depend on the reactive and dissipative parts of the electronic Green's functions respectively.
Here, $\omega_{\pm}= \omega'\pm \frac{\omega}{2}$ and ${\bf g}={\rm diag}(g, -g)$.
%Following this the real and imaginary components can be organized into the following expressions
The nontrivial bias dependence enters through the Keldysh component ${\bf G}_0^k(\omega)={\bf G}_0^{<}(\omega) +{\bf G}_0^{>}(\omega)$.
Here ${\bf G}_0^{r,a}(\omega)= \big[\omega {\bf I} \!-\! {\bf H}_{QD} \!-\! {\bf \Sigma}^{r,a}(\omega)\big]^{-1}$ and
${\bf G}_0^{</>}(\omega)= {\bf G}_0^{r}(\omega) {\bf \Sigma}^{</>}(\omega) {\bf G}_0^{a}(\omega)$ follows the Keldysh equation.
${\bf \Sigma}^{r,a,</>}(\omega)= {\bf \Sigma}_L^{r,a,</>}(\omega)+{\bf \Sigma}_R^{r,a,</>}(\omega)$
are different components of the total self-energy, additive in the metallic leads,
associated with the transfer of electrons between the metals and dots.
${\bf \Sigma}_L^{r,a}(\omega)= {\rm diag}(\mp\frac{i \Gamma_L}{2},0)$, ${\bf \Sigma}_L^{<}(\omega)= {\rm diag} \big(i f_L(\omega) \Gamma_L , 0\big)$, ${\bf \Sigma}_L^{>}(\omega)= {\rm diag} \big(-i [1\!-\!f_L(\omega)] \Gamma_L , 0\big)$. In writing the components  ${\bf \Sigma}_L^{r,a}(\omega)$ we ignore the real part responsible for the renormalization of the DQDs' energies.
Similar expressions hold for the right lead self-energy, with $\Gamma_L \to \Gamma_R$ and $f_L (\omega)\to f_R(\omega)$;
$f_{\alpha}(\omega)=[e^{\beta (\omega-\mu_{\alpha})}+1]^{-1}$ where $\beta$ is the inverse temperature, identical
for the photonic baths (ports) and the fermionic leads.
For simplicity, in numerical calculations we assume the wide-band limit for the electronic leads, and
take the metal-dots coupling to be symmetric ($\Gamma_L\!=\!\Gamma_R\!=\!\Gamma)$.
The lesser $(<)$ and greater $(>)$ components of $F_{el}(\omega)$ describe inelastic processes responsible for
the exchange of energy between electrons and the cavity mode,
\bea
F_{el}^{</>} (\omega) \!\!&=&\!\!- i \int_{-\infty}^{\infty} \frac{d\omega'}{2\pi}\!{\rm Tr} \Big[{\bf g} \,{\bf G}_{0}^{</>}(\omega_+)\,{\bf g} \, {\bf G}_{0}^{>/<}(\omega_-) \Big]\quad \quad ,
%F_{el}^{>} (\omega) &=&- i \int_{-\infty}^{\infty} \frac{d\omega'}{2\pi} {\rm Tr} \Big[{\bf g} \,{\bf G}_{0}^{>}(\omega_+)\,{\bf g} \, {\bf G}_{0}^{<}(\omega_-) \Big]
\eea
with $F_{el}^< (-\omega)=F_{el}^> (\omega)$,  satisfying the detailed balance condition in equilibrium
$F_{el}^{>}(\omega)=e^{\beta \omega} F_{el}^{<}(\omega)$.

%\begin{figure}[htb]
%\centering
%\begin{tabular}{@{}c@{}}}
%\includegraphics[width=0.5\textwidth]{NDQD.pdf} &
%\includegraphics[scale=0.4]{QCL.pdf} \\
%\end{tabular}
%\caption{Two dimensional plot of $|t(\omega)|^2$ as a function of $\omega$ and detuning $\epsilon$ for the N-DQD model
%(left $2\times 2$ panels, also note the $log_{10}$ scale)
%and the QC model (right $2\times 2$ panels).
%Parameters are same as in Fig. \ref{trans_omega}.}
%\end{figure}

%\begin{figure}[h]
%\begin{center}
%\begin{array}{cc}
%\includegraphics[width=0.23\textwidth]{NDQD.pdf}&
%\includegraphics[width=0.23\textwidth]{QCL.pdf}
%\end{array}
%\end{center}
%\caption{Figure caption}
%\label{pics:blablabla}
%\end{figure}

%\begin{table*}[h]
%\begin{tabular}{cc}
%\includegraphics[height=2.5in]{NDQD.pdf} &
%\includegraphics[height=2.5in]{QCL.pdf} \\
%\end{tabular}
%\caption{Two dimensional plot of $|t(\omega)|^2$ as a function of $\omega$ and detuning $\epsilon$ for the N-DQD model
%(left $2\times 2$ panels, also note the $log_{10}$ scale)
%and the QC model (right $2\times 2$ panels).
%Parameters are same as in Fig. \ref{trans_omega}.}
%\end{table*}

%===========================================================================================

%To understand the 
We plot the real and imaginary components of $F_{el}^r(\omega_c)$ as a function of bias difference 
in Figs.~\ref{trans_detu}(a,b). Close to equilibrium $(\Delta \mu < \omega_c)$,  both components are negative and the
electronic system acts a dissipative bath.
In contrast, far from equilibrium $(\Delta \mu > \omega_c)$, $F_{el}''(\omega_c)$ saturates to a positive value,
a necessary condition for observing gain.
By further increasing the number of DQDs, $F_{el}^{''}(\omega_c)$ approaches $\kappa$, to yield large gain (see
Fig. \ref{trans_omega}).
Note that even for a single DQD $(N\!=\!1)$, a careful tuning of parameters allows for an enhancement of
$F_{el}^{''}(\omega)$, thus photon emission. This could be achieved by:
(i) increasing the light-matter coupling strength $g$, as $F_{el}^{''}(\omega)$ scales with $g^2$,
(ii) tuning the dot-lead hybridization $\Gamma$. For weak $\Gamma$, the dwelling time of tunnelling electrons
in the dots is long ($\sim 1/\Gamma$), resulting in an effectively-strong electron-photon interaction.
(iii) As demonstrated in Figs.~\ref{trans_detu}(c-d),
by adjusting both level-detuning $\epsilon$ and $\Gamma$ we can increase $F_{el}^{''}(\omega_c)$ considerably.
 We exemplify the dependence of gain on detuning in Fig.~\ref{avg_photon}(a). %, and later on in Fig. \ref{cascade}.
Both peak and dip in the transmission, corresponding to photon emission and absorption events, respectively,
are amplified upon increasing the number of replicas $N$.
The corresponding plot for $F_{el}^{'/''}(\omega_c)$, for $N=4$, is shown in Fig.~\ref{trans_detu}(d).

The mean photon number in the cavity is another relevant observable
$\langle \hat{n}_c \rangle= \langle \hat{a}^{\dagger} \hat{a} \rangle= i \int_{-\infty}^{\infty} \frac{d\omega}{2\pi} D^{<}(\omega)$.
In the present low-temperature limit, $ \beta \, \omega_c\gg1$, it follows
% $n_{ph}(\omega_c)\!=\!0$, it is
%
\be
\langle \hat {n}_c \rangle \!=\!  \frac{i \, N\,F_{el}^{<}(\omega_c)}{2 \, \big[ \kappa \!-\! N\,F_{el}^{''}(\omega_c) \big]}.
\ee
Therefore, it is again the competition between the charge density response and photon losses to the ports
which determines the cavity photon number, see Fig.~\ref{avg_photon}(b). 
For $N=4$, giant photon number is observed, correlated with the associated high gain in the transmission function.

{\it Quantum Cascade Model.-}
We next examine the cascade architecture,  see Fig.~\ref{model-scheme}(b).
Here, multiple single-level quantum dots are sandwiched between source and drain leads.
%The Hamiltonian of the system therefore consists of $M$ sites with fixed detuning $\epsilon$ and hopping $t$ between nearest neighbours.
The transmission is determined by Eq. (\ref{eq:tdef}) with the electronic self-energy  (\ref{eq:real-im}).
In this case, ${\bf G_0}$ and ${\bf \Sigma}$ are $M\times M$ matrices made from the corresponding non-interacting dot Hamiltonian,
and ${\bf g}=diag(g_1, g_2, \cdots, g_M)$. % with $M$ the number of QDs. 
In simulations we used $g_1=-g_2$, and $g_{j>2}=0$, 
$\epsilon=\epsilon_{j+1}-\epsilon_j$,
to allow a clear comparison with the N-DQD model \cite{commentCas}.
It should be emphasized that in contrast to the N-DQD construction, % of Fig.~\ref{model-scheme}(a),
the self-energy $F_{el}^r(\omega)$ for the cascade model shows a non-monotonic behavior with $M$.

Our results are summarized in Fig.~\ref{cascade},
presenting significant photon amplification in the N-DQD and the QC models as a function of incoming photon
$\omega$ and detuning $\epsilon$, for different number of composite units.
%Here we couple only the first site with the cavity mode.
The N-DQD setup allows us to consistently enhance transmission with $N$, up to {\it five} orders of magnitude when $N=4$. Note, we plot here
${\rm log}_{10}|t(\omega)|$.
%(note that we are limited to describe situations with $\kappa>...$ DVIRA add comment.
The QC model shows a moderate enhancement, as we explain next.
For $M=1$, the QC model includes a single dot connected to metal leads, resulting in no optical gain
$|t(\omega)|^2 \le 1$, as the system lack a resonance condition. This can be proved by showing that
$F_{el}^{''}(\omega)\big{|}_{M\!=\!1} <0$ even far-from equilibrium \cite{commM1} . 
We also observe in Fig.~\ref{cascade} that $M=3$ operates better than $M=2$, as in the former {\it two} resonant transitions contribute to photon emission around the cavity frequency.
In contrast, $M=4$ (and other even-valued QC setups) do not support degenerate transitions, thus transmission amplitude drops
down to the $M=2$ case. By carefully tuning the QC Hamiltonian one could engineer several resonant transitions, to receive significant amplification.

%engineering off-diagonal coupling could
%prove highly beneficial as this would open up more resonant transitions

%--------------------
{\it Conclusion.-}
We described a fundamental mechanism for optical amplification,
by using large-scale hybrid quantum systems.
Gain in the cavity transmission is explained via a sum rule for the transmission function, and it is
achieved by counteracting the cavity decay rate to the ports  with the
gain-medium induced self-energy, the imaginary part of the charge density response function.
This cancellation is in effect only far-from-equilibrium.
We elaborated on this principle by testing two types of gain media:
For an N-DQD setup, the extensive scaling of the electronic self-energy renders a direct route
for realizing giant amplification in photon gain.
For the quantum cascade model %, the the self-energy does not grow monotonically with the number of dots, yet
gain can be enhanced when the Hamiltonian supports degenerate transitions. 
Our theory approaches the lasing threshold $F_{el}'' = \kappa$ from below, as we are limited to the regime
$F_{el}^{''} < \kappa$. Future work will involve investigations of the above-threshold regime, masing action and photon statistics. 

%======================================================================================
We acknowledge helpful discussions with J. Petta, T. Kontos, Y. Liu and M. Schiro.
B.K.A and D.S. were supported by the Natural Sciences and Engineering Research Council of Canada,
the Canada Research Chair Program, and the Centre for Quantum Information and Quantum Control (CQIQC)
at the University of Toronto.
M.K. thanks the hospitality of the Chemical Physics Theory Group at the Department of Chemistry at
the University of Toronto and the Initiative for the Theoretical Sciences (ITS) $-$ City University of
 New York Graduate Center, where several interesting discussions took place during this work.
He also gratefully acknowledges support from the Professional Staff Congress City University of New York award No.68193-0046. S. M. gratefully acknowledges the
support of the National Science Foundation (NSF) through Grant No. CHE-1361516.

%=================================================
%\bibliographystyle{apsrev}
%\bibliography{references}

\begin{thebibliography}{99}

\bibitem{RevNori}
%Hybrid quantum circuits: Superconducting circuits interacting with other quantum systems
Z.-L. Xiang, S. Ashhab, J. Q. You, and F. Nori,
Rev. Mod. Phys. {\bf 85}, 623 (2013).
%http://journals.aps.org/rmp/abstract/10.1103/RevModPhys.85.623

\bibitem{nori_sim}
% title = {Quantum simulation},
I. M. Georgescu, S. Ashhab, and F. Nori,
Rev. Mod. Phys. {\bf 86}, 153 (2014).
 % doi = {10.1103/RevModPhys.86.153},
%http://link.aps.org/doi/10.1103/RevModPhys.86.153

\bibitem{nori_simulator}
% title = {Quantum simulation},
I. Buluta and F. Nori, Science \textbf{326}, 108 (2009).
 % doi = {10.1103/RevModPhys.86.153},
%http://link.aps.org/doi/10.1103/RevModPhys.86.153


\bibitem{PNAS1}
%Quantum technologies with hybrid systems
G Kurizkia, P. Bertetb, Y. Kubob, K. Molmerc, D. Petrosyand, P. Rablf, and J. Schmiedmayer,
Proc. Natl. Acad. Sci. U.S.A. {\bf 112}, 3866 (2015).
% http://www.pnas.org/content/112/13/3866.abstract

\bibitem{Hincks_control}
%Controlling Quantum Devices with Nonlinear Hardware
I. N. Hincks, C. E. Granade, T. W. Borneman, and D. G. Cory,
Phys. Rev. Applied {\bf 4}, 024012 (2015).
% http://journals.aps.org/prapplied/abstract/10.1103/PhysRevApplied.4.024012


\bibitem{KLH2015}
K. Le Hur, L. Henriet, A. Petrescu, K. Plekhnov, G. Roux, and M. Schiro,
%Many-Body Quantum Electrodynamics Networks: Non-Equilibrium Condensed Matter Physics with Light,
arXiv:1505.00167v2.
% Review, some discussion of relevant setups.


\bibitem{nat_phys_rev}
A. A. Houck, H. E. Tureci, and J. Koch,
Nat. Phys. {\bf 8}, 292 (2012).
%   On-chip quantum simulation with superconducting circuits
% http://www.nature.com/nphys/journal/v8/n4/pdf/nphys2251.pdf

\bibitem{underwood_PhysRevA.86.023837}
%Low-disorder microwave cavity lattices for quantum simulation with photons
D. L. Underwood, W. E. Shanks, J. Koch, and A. A. Houck,
Phys. Rev. A {\bf 86}, 023837 (2012).
% http://journals.aps.org/pra/abstract/10.1103/PhysRevA.86.023837
% coupled-transmission-line resonators

\bibitem{sskj}
S. Schmidt and J. Koch,
Annalen der Physik {\bf 525}, 395 (2013).
% Circuit QED lattices: towards quantum simulation with superconducting circuits
% http://onlinelibrary.wiley.com/doi/10.1002/andp.201200261/abstract
%Jaynes-Cummings model

\bibitem{esslinger10}
K. Baumann, C.  Guerlin, F.  Brennecke and T. Esslinger, Nature {\bf 464}, 1301 (2010).
% Dicke Quantum Phase Transition with a Superfluid Gas in an Optical Cavity
% http://www.nature.com/nature/journal/v464/n7293/full/nature09009.html
% Dicke model with Bose einstein condensate

\bibitem{kulkarniprl}
%Cavity-Mediated Near-Critical Dissipative Dynamics of a Driven Condensate
M. Kulkarni,  B. \"Oztop, and H. E. T\"ureci,
Phys. Rev. Lett. {\bf 111}, 220408 (2013).

\bibitem{Krinner2015}
S.   Krinner, D. Stadler, D. Husmann, J.-P. Brantut, and T. Esslinger,
Nature {\bf 517}, 64 (2015).
% Observation of quantized conductance in neutral matter
%  www.nature.com/nature/journal/v517/n7532/full/nature14049.html
% neutral atoms

\bibitem{ess_science}
R. Mottl, F. Brennecke, K. Baumann, R. Landig, T. Donner, and T. Esslinger,
Science \textbf{336}, 1570 (2012).
% Observation of quantized conductance in neutral matter
%  www.nature.com/nature/journal/v517/n7532/full/nature14049.html
% neutral atoms


\bibitem{kulkarniprx}
 % title = {Photon-Mediated Interactions: A Scalable Tool to Create and Sustain Entangled States of $N$ Atoms},^M
C. Aron, M. Kulkarni, and H. E. T\"ureci,
Phys. Rev. X, {\bf 6} 011032 (2016).
%http://journals.aps.org/prx/abstract/10.1103/PhysRevX.6.011032

\bibitem{kulberkeley}
 M. E. Schwartz, L. Martin,  E. Flurin,  C. Aron, M. Kulkarni,  H. E. Tureci, and I. Siddiqi,
arXiv:1511.00702.

\bibitem{kulpra}
C. Aron, M. Kulkarni, and H. E. T\"ureci,
%Comments: 5 pages, 3 figures. Published version
Phys. Rev. A \textbf{90}, 062305 (2014).



%=========================================

\bibitem{kontoskondo}
%  title={Coupling a Quantum Dot, Fermionic Leads, and a Microwave Cavity on a Chip},^M
M. R. Delbecq, V. Schmitt, F. D. Parmentier, N. Roch, J. J. Viennot, G. Fève, B. Huard, C. Mora, A. Cottet, and T. Kontos,
Phys. Rev. Lett. {\bf 107}, 256804 (2011).
% http://journals.aps.org/prl/abstract/10.1103/PhysRevLett.107.256804
%  a hybrid architecture consisting of a quantum dot circuit coupled to a single mode of the electromagnetic field.


\bibitem{kontosnatcom}
%Photon-mediated interaction between distant quantum dot circuits
M. R. Delbecq, L. E. Bruhat, J. J. Viennot, S. Datta, A. Cottet, and T. Kontos,
Nat. comm. {\bf 4}, 1400 (2013).
% http://www.nature.com/ncomms/journal/v4/n1/abs/ncomms2407.html
% couple two quantum dot circuits.

\bibitem{Viennot2013}
%Out-of-equilibrium charge dynamics in a hybrid circuit quantum electrodynamics architecture
 J. J. Viennot, M. R. Delbecq, M. C. Dartiailh, A. Cottet, and T. Kontos,
Phys. Rev. B {\bf 89}, 165404 (2014).
%http://journals.aps.org/prb/abstract/10.1103/PhysRevB.89.165404
% experiment


\bibitem{Petersson2012}
K. D. Petersson, L. W. McFaul, M. D. Schroer, M. Jung, J. M. Taylor, A. A. Houck, and J. R. Petta, Nature {\bf 490}, 380 (2012).
% http://www.nature.com/nature/journal/v490/n7420/full/nature11559.html
% Circuit quantum electrodynamics with a spin qubit

\bibitem{petta2014}
%Photon Emission from a Cavity-Coupled Double Quantum Dot
Y.-Y. Liu, K. D. Petersson, J. Stehlik, J. M. Taylor, and J. R. Petta,
Phys. Rev. Lett. {\bf 113}, 036801 (2014).
% http://journals.aps.org/prl/abstract/10.1103/PhysRevLett.113.036801


\bibitem{nori1}
S. Ashhab, J.R. Johansson, A.M. Zagoskin, and F. Nori,
%Single-artificial-atom lasing using a voltage-biased superconducting charge qubit
New J. Phys. \textbf{11}, 023030 (2009).
%Photon Emission from a Cavity-Coupled Double Quantum Dot
%Y.-Y. Liu, K. D. Petersson, J. Stehlik, J. M. Taylor, and J. R. Petta,
%Phys. Rev. Lett. {\bf 113}, 036801 (2014).
% http://journals.aps.org/prl/abstract/10.1103/PhysRevLett.113.036801



\bibitem{Frey2012}
%Dipole Coupling of a Double Quantum Dot to a Microwave Resonator
T. Frey, P. J. Leek, M. Beck, A. Blais, T. Ihn, K. Ensslin, and A. Wallraff,
Phys. Rev. Lett. {\bf 108}, 046807 (2012).
% hybrid solid-state quantum device
% http://journals.aps.org/prl/pdf/10.1103/PhysRevLett.108.046807


\bibitem{enslinprl}
%  title = {Microwave Emission from Hybridized States in a Semiconductor Charge Qubit},^M
A. Stockklauser, V. F. Maisi, J. Basset, K. Cujia, C. Reichl, W. Wegscheider, T. Ihn, A. Wallraff, and K. Ensslin,
Phys. Rev. Lett. {\bf 115}, 046802 (2015).
% http://dx.doi.org/10.1103/PhysRevLett.115.046802
% microwave radiation emitted from a biased double quantum dot
% READ!


\bibitem{Toida2013}
%Vacuum Rabi Splitting in a Semiconductor Circuit QED System
H. Toida, T. Nakajima, and S. Komiyama,
Phys. Rev. Lett. {\bf 110}, 066802 (2013).
% http://journals.aps.org/prl/abstract/10.1103/PhysRevLett.110.066802
%  strong coupling regime [(g,γ,κ)≈(30,25,8.0)  MHz].

\bibitem{Kulkarni2014}
%Cavity-coupled double-quantum dot at finite bias: Analogy with lasers and beyond
M. Kulkarni, O. Cotlet, and H. E. T\"ureci,
Phys. Rev. B {\bf 90}, 125402 (2014).
% comparison to experiments- but not the actual experiments.
% http://journals.aps.org/prb/abstract/10.1103/PhysRevB.90.125402


\bibitem{Deng2013}
%Charge Number Dependence of the Dephasing Rates of a Graphene Double Quantum Dot in a Circuit QED Architecture
G.-W. Deng, D. Wei, J. R. Johansson, M.-L. Zhang, S.-X. Li, H.-O. Li, G. Cao, M. Xiao, T. Tu, G.-C. Guo, H.-W. Jiang, F. Nori, and G.-P. Guo,
Phys. Rev. Lett. {\bf 115}, 126804 (2015).
%  http://journals.aps.org/prl/abstract/10.1103/PhysRevLett.115.126804
%

\bibitem{kontosprx}
%  title = {Cavity Photons as a Probe for Charge Relaxation Resistance and Photon Emission in a Quantum Dot Coupled to Normal and Superconducting Continua},
L. E. Bruhat, J. J. Viennot, M. C. Dartiailh, M. M. Desjardins, T. Kontos, and A. Cottet,
Phys. Rev. X {\bf 6}, 021014 (2016).
% http://journals.aps.org/prx/abstract/10.1103/PhysRevX.6.021014
% Microwave cavities as a probe for the dynamics of charge transfer between a discrete electronic level and fermionic continua.


\bibitem{gpg}
G.-W. Deng, L. Henriet, D. Wei, S.-X. Li, H.-O. Li, G. Cao, M. Xiao, G.-C. Guo, M. Schiro, K. Le Hur, and G.-P. Guo,
arXiv:1509.06141.


\bibitem{Vavilov-fcs}
%Full counting statistics of photons emitted by a double quantum dot
C. Xu and M. G. Vavilov, Phys. Rev. B \textbf{88}, 195307 (2013).


\bibitem{Schon_noise} J. Jin, M. Marthaler,  P.Q. Jin, D. Golubev, G. Sch\"on, 
%Noise spectrum of a quantum dot-resonator lasing circui
New J. Phys {\bf 15}, 025044 (2013).


%-----------new paper-BJ
\bibitem{spin-kontos}
%Coherent coupling of a single spin tomicrowave cavity photons
J. J. Viennot, M. C. Dartiailh, A. Cottet, and T. Kontos, Science \textbf{349}, 408 (2015).

\bibitem{Nori-bistable} 
N. Lambert, F. Nori, and C. Flindt, Phys. Rev. Lett {\bf 115}, 216803 (2015).



\bibitem{ss2}
%Microwave quantum optics and electron transport through a metallic dot strongly coupled to a transmission line cavity
C. Bergenfeldt and P. Samuelsson
Phys. Rev. B \textbf{85}, 045446 (2012).


\bibitem{ss1}
%Non-local transport properties of nanoscale conductor-microwave cavity systems
C. Bergenfeldt and P. Samuelsson
Phys. Rev. B \textbf{87}, 195427 (2013).


%%%%%%

\bibitem{tb1}
L. D. Contreras-Pulido, C. Emary, T. Brandes, and R. Aguado, New J. Phys. \textbf{15}, 095008 (2013).



%---------------------------
%  Theory


\bibitem{ssn1}
N. Lambert, C. Flindt, and F. Nori,
Europhys. Lett. \textbf{103}, 17005 (2013).
%----------------------------------

\bibitem{tk2011}
%Mesoscopic admittance of a double quantum dot
A. Cottet, C. Mora, and T.  Kontos,  Phys. Rev. B {\bf 83}, 121311 (2011).

\bibitem{Schon}
%Lasing and transport in a multilevel double quantum dot system coupled to a microwave oscillator
C. Karlewski, A. Heimes, and G. Sch\"on,
Phys. Rev. B {\bf 93} 045314 (2016).
% http://journals.aps.org/prb/abstract/10.1103/PhysRevB.93.045314

%===============================
\bibitem{pettamaserscience}
%  title = {Semiconductor double quantum dot micromaser},^M
Y.-Y. Liu, J. Stehlik,  C. Eichler,  M. J. Gullans,  J. M. Taylor, and J. R. Petta,
Science {\bf 347}, 285 (2015).

%===============================
\bibitem{qcl}
%  title = {Semiconductor double quantum dot micromaser},^M
J. Faist,  F. Capasso, D. L. Sivco, C. Sirtori, A. L. Hutchinson, A. Y. Cho,
Science {\bf 264},  5158 (1994).

%==================================

% input output expressions
\bibitem{input_output_Clerk}
%Introduction to quantum noise, measurement, and amplification
A. A. Clerk, M. H. Devoret, S. M. Girvin, F. Marquardt, and R. J. Schoelkopf, Rev. Mod. Phys. {\bf 82}, 1155 (2010).


\bibitem{marco14}
%Tunable hybrid quantum electrodynamics from nonlinear electron transport
M. Schiro and K. Le Hur,
Phys. Rev. B {\bf 89}, 195127 (2014).
% http://journals.aps.org/prb/abstract/10.1103/PhysRevB.89.195127

\bibitem{simon16}
%Out-of-equilibrium quantum dot coupled to a microwave cavity
O. Dmytruk, M. Trif, C. Mora, and P. Simon,
Phys. Rev. B {\bf 93}, 075425 (2016).
% http://journals.aps.org/prb/abstract/10.1103/PhysRevB.93.075425
%================================

\bibitem{comment1}
In the wide-band limit we define $\kappa= 2 \pi \rho |\nu|^2$,
where $\rho$ is the photon bath density of states and $\nu$ being the average coupling between the cavity and the bath modes.


\bibitem{comment-coup}
The linear scaling of $F^r_{el}$ with the number of electronic units in the N-DQD model is valid in
the present weak coupling limit, $g/\omega_c\ll1$,
as different dissipation/response terms affect the cavity mode in an additive manner.  
%At strong couplings, saturation effects will result in more complex scaling behvior.

\bibitem{NEGF-Rammer}
J. Rammer, {\it Quantum Field Theory of Non-Equilibrium States}, Cambridge University Press, (2007).


\bibitem{longNEGF}
B. K. Agarwalla, M. Kulkarni, S. Mukamel, and D. Segal,
%Tunable photonic cavity coupled to a voltage-biased double quantum dot system: Diagrammatic NEGF approach
arXiv:1604.01811.

\bibitem{commentCas}
Though only the first two dots are coupled to the cavity mode, in the eigenenergy representation of the cascade all transitions
are coupled to the microwave mode.


\bibitem{commM1}
It can be shown that in the $M$=1 cascade model with $\Gamma_L=\Gamma_R=\Gamma$,
%
\be
\quad \quad F^{''}_{el}(\omega)\big{|}_{M\!=\!1} \!=\! g^2\! \int_{-\infty}^{\infty} \!\frac{d\omega'}{4\pi} f_n(\omega') A(\omega') \big[A(\omega'\!-\!\omega)\!-\!A(\omega'\!+\!\omega)\big] \nonumber
\ee
%
with $f_n(\omega)=\left[f_L(\omega) + f_R(\omega)\right]/2$ as
the non-equilibrium distribution function and $A(\omega)=i \left[G_0^r(\omega)-G_0^a(\omega)\right]$ the spectral function for a single-dot model. The integral of the first term is always smaller than the second term. 

\end{thebibliography}

\end{document}